\documentclass[twocolumn,groupedaddress]{revtex4}
\usepackage{amssymb}
\usepackage{amsmath}
\usepackage{graphicx}

\begin{document}

\title{Accurate sampling using Langevin dynamics}
\author{Giovanni Bussi}
\email{gbussi@ethz.ch}
\author{Michele Parrinello}
\affiliation{Computational Science, Department of Chemistry and Applied Biosciences,
ETH Z\"urich, USI Campus, Via Giuseppe Buffi 13, CH-6900 Lugano, Switzerland}
\date{\today}

\begin{abstract}
We show how to derive a simple integrator for the Langevin equation and
illustrate how it is possible to check the accuracy of the obtained
distribution on the fly, using the concept of effective energy
introduced in a recent paper [J.~Chem.~Phys.~{\bf 126}, 014101 (2007)].
Our integrator leads to correct sampling also in the difficult high-friction
limit.
We also show how these ideas can be applied in practical simulations,
using a Lennard-Jones crystal as a paradigmatic case.
\end{abstract}

\maketitle

\section{Introduction}

Langevin dynamics was first introduced in molecular simulations
to calculate the properties of mesoscopic systems~\cite{turq+77jcp}.
Here a dissipative force and a noise were added
to the Hamilton equations to model a bath of lighter particles.
The formal justification for this model can be obtained using the projection
operator techniques \cite{zwan61pr,oett98pre}.
However, it was soon realized that
Langevin dynamics can also be used as a thermostat
\cite{schn-stol78prb}, adding the dissipative forces and the noise
to the Hamiltonian dynamics to allow a molecular dynamics simulation
to explore an ensemble at a fixed temperature.
Furthermore, it has been used to sample arbitrary distribution,
for instance in the case of numerical quantum-chromodynamics \cite{batr-85prd}.

Several algorithms have been proposed for the numerical integration of
the Langevin equation, see among others Refs.~\cite{helf79bstj,green-helf81bstj,%
vang-bere82mp,brun+84cpl,%
alle-tild87book,hers98jcp,pate-ferg99cp,forb-chin01pre,skee-izag02mp,ricc-cicc03mp,scem+06jcp,
vand-cicc06cpl,mann04pre}.
Most of them were derived with the aim of producing accurate
trajectories, i.e.~dynamical properties, up to a given order.
Because of that, they usually break down when a high friction is
applied, essentially when the velocities are varying too
fast with respect to the chosen time step.
Moreover, their design is not focused
on the correctness of the ensemble generated.
A notable exception is given by the schemes derived in Ref.~\cite{mann04pre},
where the free parameters of the algorithm are
chosen so as to minimize the sampling errors.
However, none of the algorithms so far proposed
offer any
way of checking the accuracy of the sampling during a numerical simulation.
This is at variance with the numerical integration of
Hamilton's equations, where the conservation of the total
energy has been traditionally used
to this end \cite{alle-tild87book,frenk-smit02book}.
The standard approach in molecular dynamics is thus to
choose the time step by
monitoring the energy conservation in a few microcanonical runs,
then to adopt the same time step for the Langevin dynamics.
To the best of our knowledge, only in a recent paper \cite{scem+06jcp}
Scemama \emph{et al.} have shown how to correct
exactly the discretization errors in the Langevin dynamics
in the context of variational Monte Carlo, using a Metropolis procedure.
However, the poor scaling
of these accept-reject algorithms with respect to the number
of degrees of freedom prevents their application to global moves in
very large systems \cite{duan+87plb,frenk-smit02book}.

In a recent paper \cite{buss+07jcp} we introduced a
constant-temperature molecular-dynamics method. In that context,
we discussed the notion of effective energy, as a measure of sampling accuracy.
In Ref.~\cite{buss+07jcp} only one variable, the total kinetic energy,
was subject to stochastic fluctuations and the response of the
thermostat could be modeled so as to have a minimal effect on the dynamics.
Here we apply some of the ideas developed in Ref.~\cite{buss+07jcp}
to Langevin dynamics, where all degrees of freedom can be separately
controlled and the time scale over which the thermostat reacts is defined
by the friction coefficient. When used as thermostat, Langevin dynamics
can be more efficient in difficult cases, but it is more disruptive of the dynamics.
In an extension of Ref.~\cite{buss+07jcp}, we integrate Langevin
using a simple algorithm derived from a
Trotter decomposition.
The effective-energy drift allows
the sampling error to be controlled during a simulation,
and can be used in a rigorous way to perform reweighting or
accept-reject algorithms, in a scheme that turns out to be similar
to that discussed by Scemama \emph{et al.}~\cite{scem+06jcp}.
The advantage of our formulation is that, for large systems,
the effective energy can be simply checked
against long-term drifts, in the same way as the total energy has traditionally been
used to check the accuracy of microcanonical molecular dynamics.
We also show the properties of the effective energy in model
harmonic oscillators and in a realistic Lennard-Jones crystal.

\section{Theory}
\subsection{Langevin dynamics}

We consider a particle with mass $m$ subject to a potential energy $U(q)$.
The generalization to multiple degrees of freedom is straightforward.
The probability density for the canonical ensemble at an inverse temperature $\beta$ is
\begin{equation}
\label{eq-canonical}
\bar{P}(p,q)~dpdp \propto e^{-\beta\frac{p^2}{2m}}e^{-\beta U(q)}~dpdp.
\end{equation}
The canonical ensemble can be sampled through the Langevin dynamics
\begin{subequations}
\label{eq-langevin}
\begin{align}
dp(t) & = f(q(t))~dt - \gamma p(t)~dt + \sqrt{\frac{2m\gamma}{\beta}}~dW(t)
\\
dq(t) & = \frac{p(t)}{m}~dt,
\end{align}
\end{subequations}
where $f(q)=-\frac{\partial U}{\partial q}$ is the deterministic force,
$\gamma$ is the friction coefficient,
and $dW(t)$ is a Wiener noise in the Itoh convention \cite{gard03book},
normalized as
$
\langle
dW(t) dW(t')
\rangle
=
\delta(t-t')
$.
A description equivalent to the stochastic Eq.~(\ref{eq-langevin})
can be formulated in terms of the probability density,
which evolves according to the
Fokker-Planck equation \cite{gard03book,risk89book}
\begin{equation}
\label{eq-fpe}
\frac{\partial P(p,q;t)}{\partial t} = -\hat{L} P(p,q;t)
\end{equation}
where
\begin{equation}
\hat{L} = f(q)\frac{\partial}{\partial p} 
+ \frac{p}{m}\frac{\partial}{\partial q}
         - \gamma \left(
           \frac{\partial}{\partial p}p + \frac{m}{\beta}\frac{\partial^2}{\partial p^2}
         \right).
\end{equation}
The formal solution of Eq.~(\ref{eq-fpe}) at a finite time step $\Delta t$ is
\begin{equation}
P(p,q;t+\Delta t) = e^{-\Delta t\hat{L}} P(p,q;t),
\end{equation}
which however cannot be evaluated explicitly.
Notice that for Hamiltonian dynamics, $\gamma=0$,
the operator $\hat{L}$ is anti-Hermitian and
the propagator $e^{-\Delta t\hat{L}}$ is unitary.
These properties hold only for a deterministic
area-preserving dynamics.
They do not hold in a Langevin process.

\subsection{A simple integrator}

As was first recognized by Tuckerman \emph{et al.} \cite{tuck+92jcp} and, independently, by
Sexton and Weingarten \cite{sext-wein92npb}, the Trotter formula \cite{trot59pams} allows an approximated
propagator to be constructed as
\begin{equation}
\label{eq-trotter}
e^{-\Delta t \hat{L}}
\approx
\prod_{j=M}^1
e^{-\frac{\Delta t}{2} \hat{L}_j}
\prod_{k=1}^M
e^{-\frac{\Delta t}{2} \hat{L}_k}
\end{equation}
where $M$ is the number of stages in the integrator
and $\sum_j \hat{L}_j=\hat{L}$.
Since in general the $\hat{L}_j$'s do not commute among themselves,
the order in which the stages are applied is relevant,
and the splitting in Eq.~(\ref{eq-trotter}) introduces some error into the propagation.
The key point here is that
the  stages $e^{-\frac{\Delta t}{2} \hat{L}_j}$ are chosen
so that they can be integrated analytically, and
the Trotter splitting is the only source of errors.

It is natural to write $\hat{L}$  as a sum of three parts:
\begin{equation}
\hat{L} = \hat{L}_p + \hat{L}_q + \hat{L}_{\gamma}
\end{equation}
which are defined as
\begin{subequations}
\begin{align}
  \hat{L}_p  & = f(q)\frac{\partial}{\partial p}
\\
  \hat{L}_q  & =  \frac{p}{m}\frac{\partial}{\partial q}
\\
  \hat{L}_{\gamma}  & = - \gamma \left( \frac{\partial}{\partial p}p  +
\frac{m}{\beta}\frac{\partial^2}{\partial p^2}
\right).
\end{align}
\end{subequations}
Several choices are now available for the Trotter splitting.
We notice that the operators $e^{-\frac{\Delta t}{2}\hat{L}_{\gamma}}$
and $e^{-\Delta t\hat{L}_{pq}}$ leave the stationary
distribution in Eq.~(\ref{eq-canonical}) unchanged:
\begin{equation}
e^{-\frac{\Delta t}{2}\hat{L}_{\gamma}} \bar{P} = \bar{P} ~ ; ~
e^{-\Delta t\hat{L}_{pq}} \bar{P} = \bar{P}.
\end{equation}
This is due to the fact that the canonical distribution is
stationary not only with respect to $\hat{L}$ but
also with respect to $\hat{L}_{pq}=\hat{L}_p+\hat{L}_q$,
which corresponds to Hamilton propagation,
and with respect to $\hat{L}_{\gamma}$,
which introduces the combined effect of friction and noise.
Thus, even if the commutator $[\hat{L}_{pq},\hat{L}_{\gamma}]\neq 0$,
the following splitting does not introduce sampling errors,
\begin{equation}
\label{eq-first-splitting}
e^{-\Delta t\hat{L}}
\approx
e^{-\frac{\Delta t}{2}\hat{L}_{\gamma}}
e^{-\Delta t\hat{L}_{pq}}
e^{-\frac{\Delta t}{2}\hat{L}_{\gamma}},
\end{equation}
since it can be interpreted as a sequence of moves each of which
has the correct limiting distribution.
The $e^{-\frac{\Delta t}{2}\hat{L}_{\gamma}}$ move provides ergodicity
in the momenta subspace only, while the $e^{-\Delta t\hat{L}_{pq}}$ move
mixes the momenta and positions subspaces.
An integrator designed to apply the propagator in Eq.~(\ref{eq-first-splitting})
would provide an approximate trajectory
and an exact sampling, independently of $\Delta t$ and $\gamma$.
The propagator $e^{-\frac{\Delta t}{2}\hat{L}_{\gamma}}$
can be integrated analytically.
Unfortunately, the propagator $e^{-\Delta t\hat{L}_{pq}}$
cannot be integrated exactly and has to be split further.
We opt here for the simplest choice, which is the same used to obtain
the velocity Verlet algorithm:
\begin{equation}
\label{eq-splitting}
e^{-\Delta t\hat{L}}
\approx
e^{-\frac{\Delta t}{2}\hat{L}_{\gamma}}
e^{-\frac{\Delta t}{2}\hat{L}_{p}}
e^{-\Delta t\hat{L}_{q}}
e^{-\frac{\Delta t}{2}\hat{L}_{p}}
e^{-\frac{\Delta t}{2}\hat{L}_{\gamma}}.
\end{equation}
In specific cases, different decompositions of $\hat{L}_{pq}$
could be adopted.
For example, if the forces can be separated into contributions varying
on different time scales, a multiple-time-step
decomposition is expected to be more efficient~\cite{tuck+92jcp}.

Other possible choices for the Trotter splitting which are
substantially equivalent to Eq.~(\ref{eq-splitting}) can be
obtained, based on the three operators
$\hat{L}_{q}$, $\hat{L}_{p}$ and $\hat{L}_{\gamma}$.
It is worthwhile to notice that in principle
there is no need to split $\hat{L}_{p}$ and
$\hat{L}_{\gamma}$, since 
$\hat{L}_{p\gamma} = \hat{L}_{p} + \hat{L}_{\gamma}$
can be also evolved analytically.
Ricci and Ciccotti~\cite{ricc-cicc03mp} derived
two integrators  using splittings that, in our
notation, would read
$e^{-\Delta t\hat{L}}
\approx
e^{-\frac{\Delta t}{2}\hat{L}_{p\gamma}}
e^{-\Delta t\hat{L}_{q}}
e^{-\frac{\Delta t}{2}\hat{L}_{p\gamma}}
$
and
$e^{-\Delta t\hat{L}}
\approx
e^{-\frac{\Delta t}{2}\hat{L}_{q}}
e^{-\Delta t\hat{L}_{p\gamma}}
e^{-\frac{\Delta t}{2}\hat{L}_{q}}
$.
These decompositions involve a single splitting and
thus appear more accurate than Eq.~(\ref{eq-splitting}).
However, when $\gamma \Delta t$ is negligible,
they do not offer any advantage, and when
 $\gamma \Delta t$ is not negligible, they do not sample
the proper ensemble. This can be easily verified taking the limit
$\gamma \Delta t \rightarrow \infty$.
On the other hand, in our scheme the only ensemble violations
arise from the fact that for a finite $\Delta t$ the evolution
of $\hat{L}_{pq}$ is approximated.
These violations are independent of the choice of the
friction. Even the infinite friction limit can be taken
safely, as shown in Appendix~\ref{sec-high-friction-limit}.
Thus, when the sampling quality is an issue, our scheme offers
significant advantages.

The splitting in Eq.~(\ref{eq-splitting}) leads to an explicit integration scheme.
In the derivation we use the analytical propagation 
formula for $\hat{L}_{\gamma}$ which can be found in Ref.~\cite{risk89book}.
After some manipulation, the integrator is written as:
\begin{subequations}
\label{eq-integrator}
\begin{align}
\label{eq-integrator1}
p(t^+)&=c_1 p(t) + c_2 R(t)
\\
\label{eq-integrator2}
q(t+\Delta t) &= q(t) + \frac{p(t^+)}{m}\Delta t
+ \frac{f(q(t))}{m} \frac{\Delta t^2}{2}
\\
\label{eq-integrator3}
p(t^-+\Delta t) &= p(t^+) + \frac{f(q(t))+f(q(t+\Delta t))}{2} \Delta t
\\
\label{eq-integrator4}
p(t+\Delta t) &= c_1 p(t^-+\Delta t) +
 c_2 R'(t+\Delta t)
\end{align}
\end{subequations}
where $R$ and $R'$ are two independent Gaussian numbers
and the coefficients $c_1$ and $c_2$ are 
\begin{subequations}
\begin{align}
\label{eq-c1}
c_1&=e^{-\gamma \frac{\Delta t}{2}}
\\
\label{eq-c2}
c_2&=\sqrt{(1-c_1^2)\frac{m}{\beta}}.
\end{align}
\end{subequations}
Equation~(\ref{eq-c2}) fixes the
weight of the rescaling factor $c_1$ and of the
amplitude of the Gaussian number $c_2$ in such a way
that $c_1p+c_2R$ will be distributed in the same way as $p$.
Thus, Eq.~(\ref{eq-c2}) alone guarantees the correctness of the sampling.
On the other hand, Eq.~(\ref{eq-c1}) gives the relation between
the friction $\gamma$ and the rescaling factor $c_1$.

In Equation~(\ref{eq-integrator}),
the combination of the two inner stages is a velocity Verlet step,
and corresponds to the approximate propagation of 
$e^{-\Delta t\hat{L}_{pq}}$.
The first and last stages represent the action of the thermostat,
i.e.~the exact propagation of $e^{-\frac{\Delta}{2} t\hat{L}_{\gamma}}$.
We denote as $p(t^+)$ and $p(t^-)$
the momenta immediately after and immediately before the action of the thermostat.
We also observe that
the first and last stages can be merged as 
$
p(t^++\Delta t) = c_1^2p(t^-+\Delta t)+c_2\sqrt{c_1^2+1}R(t+\Delta t)
$ so that one Gaussian random number per degree of freedom is required at each step.
This allows the simulation to speed up when the calculation
of the deterministic forces is particularly cheap and the
generation of the Gaussian random numbers becomes computationally relevant.
If one is interested in the values of the momenta at time $t$, i.e. synchronized
with the positions, they can be reconstructed afterwards.

\subsection{Control of sampling errors}

We now use the concept of effective energy $\tilde{H}$ introduced in Ref.~\cite{buss+07jcp}
to control the accuracy of the sampling. For clarity
we repeat here some of the notions already presented there.

Our goal is to generate a sequence of points $x_i=(p_i,q_i)$ in the phase-space,
so that a time average can be used in place of the ensemble average~\cite{alle-tild87book}.
Usually, in molecular dynamics simulations this sampling is
approximate, due to the finite-time-step errors.
On the other hand, in a Monte Carlo simulation the moves are
accepted or refused in such a way that the exact distribution is enforced.
Here, we interpret a stochastic molecular dynamics as
a highly efficient Monte Carlo where all the moves are accepted.
We define $M(x_{i+1}\leftarrow x_i)~dx_{i+1}$ the distribution probability of
the point $x_{i+1}$ to be chosen as the next point, given that the present
point is $x_i$. We also define the conjugate point $x^*=(-p,q)$,
which is obtained by inverting the momentum, and satisfies $\bar{P}(x)=\bar{P}(x^*)$.
If Equation~(\ref{eq-langevin}) was integrated exactly,
then the detailed balance \cite{gard03book} would be satisfied, i.e.,
$M(x_{i+1}\leftarrow x_i)\bar{P}(x_i)=M(x_i^*\leftarrow x_{i+1}^*)\bar{P}(x_{i+1}^*)$.
However, this is not true when a finite time step is used.
Thus, we introduce a weight $w_i$ associated to the point $x_i$,
which evolves as
\begin{equation}
\frac{w_{i+1}}{w_i} = \frac{M(x^*_i\leftarrow x^*_{i+1})\bar{P}(x_{i+1})}{M(x_{i+1}\leftarrow x_i)\bar{P}(x_i)}.
\end{equation}
The same information can be expressed in terms of an effective energy,
defined as $\tilde{H}_i=-\frac{1}{\beta}\log w_i$, which evolves according to
\begin{equation}
\label{eq-effective-energy-drift1}
\tilde{H}_{i+1} - \tilde{H}_i = -\frac{1}{\beta} \log \left(
\frac{M(x^*_i\leftarrow x^*_{i+1})}{M(x_{i+1}\leftarrow x_i)}
\right)
+ H(x_{i+1})-H(x_i).
\end{equation}
We now proceed into an explicit derivation of the terms needed.

In standard hybrid Monte Carlo, the trial moves
are generated using an area-preserving scheme, so that
$M(x^*_i\leftarrow x^*_{i+1})=M(x_{i+1}\leftarrow x_i)$.
Thus, the effective energy $\tilde{H}$ reduces to the Hamiltonian
$H$. However, the Langevin equation is explicitly non-area-preserving,
and an additional contribution due to phase-space compression has to be evaluated.
We now calculate it explicitly for the integrator in Eq.~(\ref{eq-integrator}).
In Ref.~\cite{buss+07jcp} we used the fact that the thermostat moves are designed
so as to satisfy detailed balance.  We present here
a more general way of evaluating this contribution that can be straightforwardly
applied to other integrators.

We recall that the random numbers $R$ and $R'$ are drawn
from a Gaussian distribution, i.e.
\begin{equation}
P(R,R')~dRdR' =
\frac{1}{2\pi}
e^{-{\frac{R^2}{2}}}e^{-\frac{R'^2}{2}}
~dRdR'.
\end{equation}
We notice that given the starting point $x_i=(p_i,q_i)$
and the ending point $x_{i+1}=(p_{i+1},q_{i+1})$
the value of $R$ and $R'$ can be determined
solving Eqs.~(\ref{eq-integrator})
with respect to $R$ and $R'$:
\begin{subequations}
\begin{align}
R&=
(q_{i+1}-q_i)\frac{m}{c_2\Delta t} - \frac{f(q_i)\Delta t}{2c_2}
-\frac{c_1}{c_2}p_i
\\
R'&=
-(q_{i+1}-q_i)\frac{c_1m}{c_2\Delta t} - \frac{c_1f(q_{i+1})\Delta t}{2c_2}
+\frac{1}{c_2}p_{i+1}
\end{align}
\end{subequations}
where we have identified the sequence index $i$ with the time $t$
and the sequence index $i+1$ with the time $t+\Delta t$.
Now, changing the variables from $(R,R')$ to $(q_{i+1},p_{i+1})$
one obtains the following expression for the transition probability:
\begin{multline}
M\Big((q_{i+1},p_{i+1})\leftarrow(q_i,p_i)\Big) =
\\
=
\frac{m}{2\pi c_2^2\Delta t}
\exp\Bigg(
-\frac{1}{2c_2^2}\Big((q_{i+1}-q_i)\frac{m}{\Delta t} - \frac{f(q_i)\Delta t}{2}
-c_1p_i\Big)^2
\\
-\frac{1}{2c_2^2}\Big((q_{i+1}-q_i)\frac{mc_1}{\Delta t} + \frac{f(q_{i+1})c_1\Delta t}{2}
-p_{i+1}\Big)^2
\Bigg).
\end{multline}
At this stage we know the probability for the forward move
$M((q_{i+1},p_{i+1})\leftarrow(q_i,p_i))$. With a similar procedure
we can find the probability for the backward move,
$M((q_i,-p_i)\leftarrow(q_{i+1},-p_{i+1}))$, and,
with some further manipulation, the contribution of the phase-space compression to the effective energy:
\begin{multline}
\label{eq-effective-energy-drift2}
-\frac{1}{\beta} \log \left(
\frac{M(x^*_i\leftarrow x^*_{i+1})}{M(x_{i+1}\leftarrow x_i)}
\right) 
= -\left(
\frac{p_{i+1}^2}{2m}
-
\frac{p_{i}^2}{2m}
\right)
\\
+
(q_{i+1}-q_i)\frac{f_{i+1}+f_i}{2}
+
\frac{\Delta t^2}{8m} \left(f(q_{i+1})^2 - f(q_{i})^2
\right).
\end{multline}

For this derivation it is crucial that the change of variables
be well defined. Since we have two noise terms
$(R,R')$ and two variables $(q_{i+1},p_{i+1})$, we have to require
the Jacobian of the transformation to be different from zero. For integrators
using only one noise term, 
it is not obvious that, given
the forward trajectory, the backward trajectory is
possible.
If the backward trajectory is possible, then
the effective-energy drift depends on the ratio between the 
forward and backward probabilities
and gives a quantitative measure of the violation of detailed
balance. If the backward trajectory is not possible,
then the integrator cannot satisfy detailed balance.
As an example, the second integrator introduced by Ricci and Ciccotti~\cite{ricc-cicc03mp}
cannot satisfy detailed balance, as was already pointed out by
Scemama \emph{et al.}~\cite{scem+06jcp}.
On the other hand, the  modification described in Ref.~\cite{scem+06jcp} can satisfy detailed balance.
It is interesting that in Ref.~\cite{scem+06jcp} the authors are using the usual formulation
of detailed balance, which leads to the need for an explicit inversion of the
sign of the velocities. We use a more general formulation of detailed balance~\cite{gard03book}
in which velocities are considered as odd variables and their inversion
after an accepted step is not required.
One could also object that
the condition of detailed balance is not strictly
necessary \cite{mano-deem99jcp}. However, it appears to us that
detailed balance is the only way to enforce or check a distribution in a local manner,
i.e., using only information about the present point $x_i$, the next point $x_{i+1}$,
and their conjugated points $x_i^*$ and $x_{i+1}^*$.

Equations~(\ref{eq-effective-energy-drift1}) and~(\ref{eq-effective-energy-drift2})
can be combined, giving a final expression for the effective-energy increment as
\begin{equation}
\label{eq-deltahtilde}
\Delta
\tilde{H}
=
\Delta q
\left(
\frac{f(q_i)+f(q_{i+1})}{2}
\right)
+
\Delta U
+
\frac{\Delta t^2}{8m} \Delta (f^2).
\end{equation}
From Equation~(\ref{eq-deltahtilde}) it is easy to see 
that when $\Delta t$ is small enough the effective energy is approximately constant,
since the first and second terms tend to compensate each other
and the third term vanishes on the order of $\Delta t^2$.
We also notice that the third term in Eq.~(\ref{eq-deltahtilde})
is an exact differential. Thus it contributes to the fluctuations of the effective energy
but not to its drift.

We notice that the increment of the effective energy in Eq.~(\ref{eq-deltahtilde})
is exactly equal to the difference of the total energy before and after the
velocity Verlet step, as in the case of the scaling procedure described
in Ref.~\cite{buss+07jcp}.
In fact, also here we can think of our dynamics
as composed of a combination of two steps: one, described
by the operator $e^{-\frac{\Delta t}{2}\hat{L}_{\gamma}}$, which exactly satisfied detailed
balance; the other, which is the velocity Verlet step, is symplectic
but does not exactly conserve the energy. Only the violations arising
from the latter are accumulated into the effective energy.
Thus, in practice, $\tilde{H}$ can be calculated simply by summing
the increments of the total energy due to the Verlet, discarding the increments
due to the thermostat. Alternatively, it can be obtained by subtracting from the total
energy the sum of all its increments due to the thermostat.
The same procedure can be applied directly also in the
case of the Peters thermostat \cite{pete04el},
based on dissipative particle dynamics \cite{sodd+03pre},
where $e^{-\frac{\Delta t}{2}\hat{L}_{\gamma}}$ is substituted
by a rescaling of the relative velocity of neighboring particles,
the only condition being the fact that the rescalings are
performed in a way that analytically preserves the target ensemble.

The effective energy can be calculated on the fly and, aside from
numerical truncation errors,
it gives a quantitative way to assess the accuracy of the calculation.
In the spirit of Ref.~\cite{wong-lian97pnas}, one can obtain
exact ensemble averages with $\langle A \rangle=\sum_iw_iA(x_i)/\sum_iw_i$.
The variation of $\tilde{H}$ on segments of trajectory
can also be used in a hybrid Monte Carlo scheme \cite{duan+87plb}, where the acceptance
is calculated as
$\min\left(1,e^{-\beta\Delta(\tilde{H})}\right)$. In this latter case,
our scheme becomes similar to that presented by Scemama \emph{et al.}~\cite{scem+06jcp}.
In a molecular dynamics context, the effective energy $\tilde{H}$
is simply monitored during the simulation. It may fluctuate, but
it should not exhibit a large systematic drift.

\section{Examples}
\label{sec-examples}

\subsection{Harmonic oscillator}

It is instructive to study the properties of the integrator in Eq.~(\ref{eq-integrator})
when it is applied to a harmonic oscillator.
We consider an energy profile
\begin{equation}
U(q) = \frac{1}{2}m\omega^2q^2.
\end{equation}
We are interested in the time evolution of the effective
energy $\tilde{H}$. It can be easily shown that for a quadratic potential
the first two terms in Eq.~(\ref{eq-deltahtilde}) cancel exactly, and only the third
term survives. Thus, the integral over the trajectory is not necessary
and the effective energy $\tilde{H}$ is a state function
\begin{equation}
\label{eq-htilde-harmonic}
\tilde{H}(p,q) = \frac{\Delta t^2}{8} m \omega^4q^2 + C
\end{equation}
where $C$ is an arbitrary constant.
The effective distribution that will be
sampled by the Langevin dynamics can be obtained analytically and is
\begin{equation}
\label{eq-harmonic-pe}
\bar{P}_e(p,q) \propto e^{-\beta\frac{p^2}{2m}-\beta\frac{\omega^2m}{2}\left(1-\frac{\omega^2\Delta t^2}{4}\right)q^2}.
\end{equation}
This solution can be normalized only if
$\Delta t < 2/\omega$. For longer time steps, the dynamics is unstable.
Although the logarithm of the distribution in Eq.~(\ref{eq-harmonic-pe})
has an expression similar to that of the so-called shadow
Hamiltonian for the harmonic oscillator~\cite{toxv94pre,jaso-shallo00pre}, our derivation of
Eq.~(\ref{eq-harmonic-pe}) is based on the stationary distribution only
and does not provide any information on the effective trajectory.

Since $\tilde{H}$ is a state function, it will not exhibit drifts.
Its square fluctuations can be obtained analytically and are equal to
\begin{equation}
\Delta \tilde{H}^2
=
\frac{1}{2\beta^2}\frac{1}{\left(\frac{4}{(\omega\Delta t)^2}-1\right)^2}.
\end{equation}
For a comparison, the fluctuations of the total energy $H$ are $1/\beta^2$.
Interestingly, the size of the fluctuations of $\tilde{H}$
depends only on the ratio between the time step and the period of the oscillator.
It is completely independent of the value
of the friction.

The properties of a $N$-dimensional oscillator can be easily
obtained by recalling that, when the dynamics is projected on the eigenmodes
of the oscillator, the coordinates evolve independently of each other.
Assuming a spectrum of $N$ frequencies $\omega_i$, the
fluctuations of $\tilde{H}$ are
\begin{equation}
\Delta \tilde{H}^2
=
\frac{1}{2\beta^2}\sum_{i=1}^N\frac{1}{\left(\frac{4}{(\omega_i\Delta t)^2}-1\right)^2}
\approx
\frac{\Delta t^4}{32\beta^2}\sum_{i=1}^N\omega_i^4.
\end{equation}
The last approximation holds when $\Delta t$ is much smaller
than the period of the fastest mode.
In this case, it is interesting to note that
to have rigorously the same accuracy,
the time step has to be chosen proportional
to $N^{-1/4}$.

\subsection{Lennard-Jones crystal}

In the harmonic oscillator the effective energy reduces to a state function and
does not exhibit drifts. In this sense, the harmonic oscillator cannot be considered as a
prototype of a real molecular system.
In this subsection we discuss the application of Langevin sampling
and of effective-energy monitoring in the context
of atomistic simulations. We use as a realistic test-case a Lennard-Jones
solid, close to the melting point.
We express all the quantities in reduced units \cite{alle-tild87book}.
We simulate a cubic box with side 19.06 containing 6912 particles arranged according to an fcc lattice,
which corresponds to a density $\rho$=0.998.
We set the temperature to $T$=0.667.
We calculate the forces using a distance cutoff of 3.
We compare simulations performed using different values for the time step $\Delta t$
and the friction $\gamma$.
All the simulations were performed using a modified version of the DL POLY code
\cite{dlpoly,dlpoly2}.

\begin{figure}
\includegraphics[clip,width=0.45\textwidth]{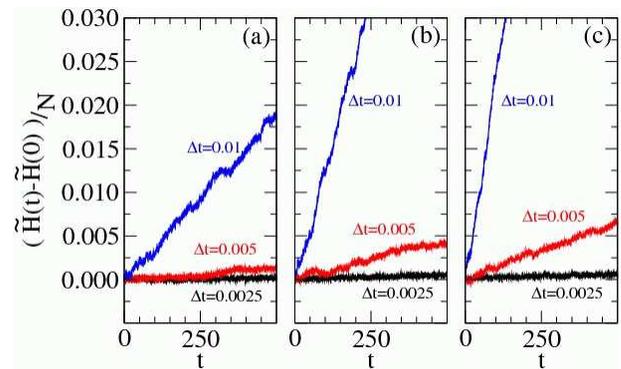}
\caption{
\label{fig-lj-drift}
(color online)
Effective-energy drifts for different choices of the friction coefficient,
respectively
$\gamma$=1 (a),
$\gamma$=5 (b) and
$\gamma$=20 (c), and different choices of the time step $\Delta t$,
as indicated. The effective energy drifts linearly, and its slope
is strongly dependent on the time step. All the quantities are in Lennard-Jones reduced units.
}
\end{figure}

\begin{figure}
\includegraphics[clip,width=0.45\textwidth]{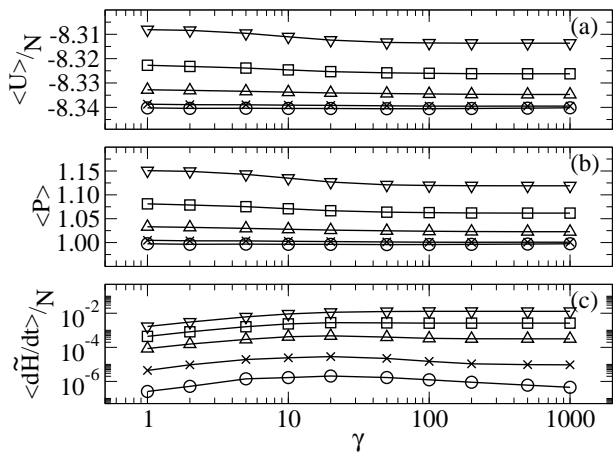}
\caption{
\label{fig-lj}
Average values of (a) the potential energy per particle, (b) the instantaneous pressure
and (c) the slope of the effective energy per particle,
plotted as functions of the friction $\gamma$.
The calculations are performed with different time steps:
$\Delta t=0.0025$ ($\circ$), $\Delta t=0.005$ ($\times$),
$\Delta t=0.01$ ($\triangle$), $\Delta t=0.015$ ($\square$) and
$\Delta t=0.02$ ($\triangledown$). All the quantities are in Lennard-Jones reduced units.
}
\end{figure}

\begin{figure}
\includegraphics[clip,width=0.45\textwidth]{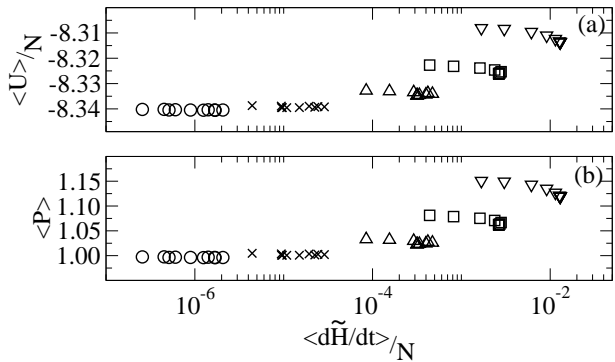}
\caption{
\label{fig-slope-err}
Average value of (a) the potential energy per particle and (b) instantaneous pressure,
plotted as functions of the slope in the effective energy drift per particle.
The calculations are performed with different time steps:
$\Delta t=0.0025$ ($\circ$), $\Delta t=0.005$ ($\times$),
$\Delta t=0.01$ ($\triangle$), $\Delta t=0.015$ ($\square$) and
$\Delta t=0.02$ ($\triangledown$). All the quantities are in Lennard-Jones reduced units.
}
\end{figure}

In Fig.~\ref{fig-lj-drift} we show a time series for the effective energy $\tilde{H}$
per particle.
The effective energy exhibits a regular drift due to the finiteness of the
integration time step, similarly to the total energy
in a microcanonical simulation.
The drift is strongly dependent on the time step, and is only slightly affected
by the choice of the friction.
In Fig.~\ref{fig-lj} we show the values obtained for the average potential
energy per particle and the average pressure for different choices of $\gamma$ and $\Delta t$,
obtained from runs of length 2500 time units.
The values are again rather independent of the choice of $\gamma$.
 This is remarkable,
considering that we are changing the friction over three orders of magnitude
and that we are working also in regimes where $\gamma\Delta t$ is not negligible.
This indicates that the errors are essentially coming from the integration
of the Hamilton equations and not from the friction itself.
In the third panel we show the average slope of the effective energy per particle,
obtained with a linear fitting.
The slope is again strongly sensitive to the time step and
only slightly dependent on the friction.

To stress the fact that the effective-energy slope is a correct indicator
of the integration errors, we show the same data in Fig.~\ref{fig-slope-err}.
There, we plot the value of the observable quantity as a function of the
slope in the effective-energy drift. The two quantities are highly correlated,
indicating that the effective-energy slope gives a realistic estimate of
the errors due to the finite time step.

\begin{figure}
\includegraphics[clip,width=0.45\textwidth]{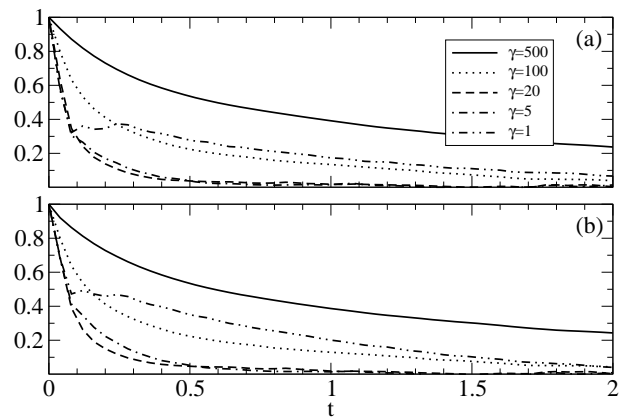}
\caption{
\label{fig-lj-auto}
Normalized autocorrelation function of (a) the potential energy and (b) the
instantaneous pressure, for different choices of the 
friction coefficient $\gamma$, as indicated.
The fastest decorrelation is observed when $\gamma$ is set to
an optimal value of 20. All the quantities are in Lennard-Jones reduced units.
}
\end{figure}

Up to now we have discussed the sampling accuracy, which measures the
systematic errors due to the finite-time-step integration.
In practical applications also the sampling efficiency, which measures the
statistical error due to the finite length of the simulation, is relevant.
The sampling efficiency depends on which specific
observable one wishes to calculate. In particular, to optimize the
efficiency, the autocorrelation
time of the quantity of interest has to be as short as possible \cite{frenk-smit02book}.
In Fig.~\ref{fig-lj-auto} we show the autocorrelation function of the 
total potential energy and of the instantaneous pressure, for different
choices of the friction $\gamma$, using a time step $\Delta t=0.0025$.
For both the considered quantities, the optimal choice
for $\gamma$ is 20. This rule is far from general,
but illustrates clearly the fact that too high a friction can spoil
the quality of the sampling since it hinders particle motion.

\section{Conclusion}

In conclusion, we have studied the properties
of a very simple integrator for the Langevin equation,
derived employing the Trotter scheme commonly
use in the derivation of multiple-time-step integrators.
Moreover, we have used the concept of effective energy,
introduced in a previous paper, to asses on the fly the
accuracy of this integrator in practical cases, ranging
from simple one-dimensional oscillators to a Lennard-Jones
crystal. Finally, we have shown how to monitor the effective energy
in practice.
Our formalism can be easily generalized to the description
of other stochastic dynamics, such as dissipative-particle
dynamics \cite{sodd+03pre}.

\appendix

\section{High friction limit}
\label{sec-high-friction-limit}
When $\gamma\rightarrow\infty$ the integrator in Eq.~(\ref{eq-integrator})
can be rewritten in terms of the position only:
\begin{equation}
q(t+\Delta t) = q(t)
+
f(q(t))\frac{\Delta t^2}{2m}
+
\Delta t
\sqrt{\frac{1}{\beta m}} R
\end{equation}
Now, defining $D=\frac{\Delta t}{2\beta m}$ this equation becomes
\begin{equation}
q(t+\Delta t) = q(t) + D \beta f(q(t)) \Delta t + \sqrt{2D\Delta t} R
\end{equation}
which is exactly the Euler integrator for the overdamped Langevin equation
\begin{equation}
dq(t) = D\beta f(q(t))~dt + \sqrt{2D}~dW(t).
\end{equation}
It is worth noting that the increment of $\tilde{H}$
as defined in Eq.~(\ref{eq-deltahtilde}) does not depend on $\gamma$, and is still
valid. In terms of $D$ it is
\begin{equation}
\Delta
\tilde{H}
=
\Delta q
\left(
\frac{f(q(t))+f(q(t+\Delta t))}{2}
\right)
+
\Delta U
+
\frac{\beta D\Delta t}{4} \Delta (f^2)
\end{equation}
which is exactly the one used to calculate the
acceptance in the smart Monte Carlo technique \cite{ross+78jcp}.

\end{document}